%% file: main_HICSS.tex
\documentclass[10pt]{article}

\usepackage{tocbasic} 

\usepackage[
    style=apa,
  ]{biblatex}
\PassOptionsToPackage{dvipsnames}{xcolor}
\usepackage[dvipsnames]{xcolor}

\usepackage[letterpaper]{geometry}
\usepackage{hicss}
\usepackage{times}             
\usepackage[none]{hyphenat}
\usepackage{xurl}               
\usepackage{url}
\usepackage{latexsym}
\usepackage{minted}
\usepackage{indentfirst}
\usepackage{graphicx}
\usepackage{float}

\usepackage{enumitem}
\setlist[itemize]{noitemsep, topsep=0pt}

\usepackage[most]{tcolorbox}
\usepackage{adjustbox}
\usepackage[linesnumbered,algo2e,ruled]{algorithm2e}
\usepackage{multirow}
\usepackage{booktabs,dcolumn,tabularx}
\usepackage{tikz}
\usepackage{pgfplots}
\pgfplotsset{compat=1.18}

\usepackage{subcaption}
\usepackage[font={small,sf,bf},labelsep=period]{caption}

\usepackage{pifont}
\usepackage{soul}
\usepackage{amsmath,amsfonts}

\usepackage{svg}

\usepackage{hyperref}
\hypersetup{
  colorlinks=true,
  linkcolor=RubineRed,
  citecolor=RubineRed,
  urlcolor=RubineRed,
  breaklinks=true
}

\graphicspath{{Figures/}}
\svgpath{{Figures/}}

\input{commands}

\renewcommand{\cite}{\textcite}

\input{title}

\author{Irdin Pekaric \\
University of Liechtenstein \\
Vaduz, Liechtenstein \\
 {\underline{irdin.pekaric@uni.li}}  \\  \And
Giovanni Apruzzese \\
University of Liechtenstein \\
Vaduz, Liechtenstein \\
 {\underline{giovanni.apruzzese@uni.li}} 
  }

\date{}

\begin{document}
\maketitle

\input{sections/0-abstract}

\subsubsection*{Keywords:}

Research, Transparency, Replicability, Reproducibility, Repository, Resources, Open Science

\input{structure}

\renewcommand*{\bibfont}{\small}
\printbibliography

\clearpage

\end{document}

%% file: commands.tex
\newcolumntype{?}{!{\vrule width 1.5pt}}

\newcommand{\textbox}[1]{
    \noindent\fbox{%
        \parbox{0.97\columnwidth}{%
            {#1}
        }%
    }
}

\newtcolorbox{cooltextbox}[1][]{%
    colback=black!5,
    colframe=black!5,
    notitle,
    sharp corners,
    borderline west={0pt}{0pt}{red!80!black},
    enhanced,
    breakable,
    left=0pt,
    right=0pt,
    top=0pt,
    bottom=0pt
    }

\newcommand\revision[1]{%
  \bgroup
  \hskip0pt\color{green!60!black}%
  #1%
  \egroup
}

%% file: title.tex
\title{``We provide our resources in a dedicated repository''\\Surveying the Transparency of HICSS publications}


%% file: sections/0-abstract.tex
\begin{abstract}

Every day, new discoveries are made by researchers from all across the globe and fields. HICSS is a flagship venue to present and discuss such scientific advances. Yet, the activities carried out for any given research can hardly be fully contained in a single document of a few pages---the ``paper.'' Indeed, any given study entails data, artifacts, or other material that is crucial to truly appreciate the contributions claimed in the corresponding paper. External repositories (e.g., GitHub) are a convenient tool to store all such resources so that future work can freely observe and build upon them---thereby improving transparency and promoting reproducibility of research as a whole.

In this work, we scrutinize the extent to which papers recently accepted to HICSS leverage such repositories to provide supplementary material. To this end, we collect all the 5579 papers included in HICSS proceedings from 2017--2024. Then, we identify those entailing either human subject research (850) or technical implementations (737), or both (147). Finally, we review their text, examining how many include a link to an external repository---and, inspect its contents. Overall, out of 2028 papers, only 3\% have a functional and publicly available repository that is usable by downstream research. We release all our tools.
\end{abstract}

%% file: structure.tex
\input{sections/1-introduction}

\input{sections/2-related}

\input{sections/3-method}

\input{sections/4-results}
\input{sections/5-discussion}

\input{sections/6-conclusions}

%% file: sections/1-introduction.tex
\section{Introduction}
\label{sec:introduction}

\noindent

The Hawaii International Conference on System Sciences (HICSS) is one of the largest outlets for scientific advances. Since 1967, HICSS represents a hub for researchers from a plethora of domains gravitating around information technology (IT) and social sciences. For instance, recently accepted works encompass digital innovation~\parencite{nijsse2023identifying}, law and regulation~\parencite{koh2024voices}, software development~\parencite{heine2024lean}, or business intelligence~\parencite{ul2019business}. The ``numbers'' of HICSS are constantly increasing (e.g., 550 papers in 2012, 635 in 2017, and 764 in 2024), indicating that, every year, more and more cutting-edge discoveries are made and are significant enough to be discussed.

\begin{figure}[!htbp]
    \vspace{-0.3cm}
    \centering
    \includegraphics[width=0.8\columnwidth]{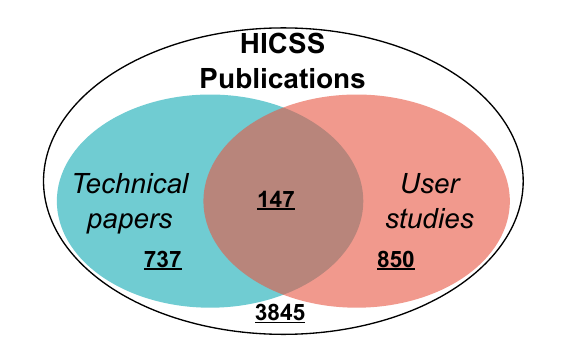}
    \vspace{-0.3cm}
    \caption{{\small \textbf{Distribution of the papers analysed in our research}} --
    \textmd{\footnotesize From 5579 papers published at HICSS since 2017, we isolate 884 technical papers and 997 user-studies papers (147 papers are shared), and examine their repositories (if any).}} 
    \label{fig:venn}
    \vspace{-3mm}
\end{figure}

The growth seen by HICSS reflects the generic trend of research~\parencite{to2023rise}. Unfortunately, despite these successes, the rising number of scientific publications may conceal various problems---among which one, in particular, stands out: the \textbf{reproducibility crisis}~\parencite{baker2016reproducibility}. Simply put, even though novel findings are published every new day, researchers are having troubles reproducing the results of prior work. This phenomenon is claimed to affect various research fields---most notably, artificial intelligence (AI), but also biology or business~\parencite{kapoor2023leakage,miyakawa2020no,aguinis2020science}. Such ``endemicity'' is problematic, since it casts doubts on the conclusions of researchers, undermining their credibility. 

Among the root causes of the reproducibility crisis, there are two culprits: the poor \textbf{replicability}~\parencite{plesser2018reproducibility} and \textbf{transparency}~\parencite{hardwicke2020empirical} of research papers. Indeed, sometimes papers may not include the information required to even attempt reproduction of the experiments described therein. For instance, some low-level details may not be provided, the implementation of an algorithm may not be disclosed, or the data to test a hypothesis may not be released. Altogether, these ``lacks'' impair reproducibility, since downstream researchers are likely to introduce some deviations in their setup when reproducing a prior work---leading to different results.

Importantly, these ``omissions'' may be due to valid reasons, such as strict page limit, data confidentiality, or proprietary/patented software~\parencite{apruzzese2023real, pekaric2021vulner}. However, while the latter reasons are unavoidable, there exist ways to address the former: external \textbf{repositories}. For instance, GitHub is a platform wherein any sort of supplementary research material (e.g., documentation, data, source code) can be uploaded~\parencite{milliken2021behavioral}, thereby representing a solution to the limited space available on written manuscripts. Still, despite the existence of such tools (which could mitigate the reproducibility crisis and improve transparency of research), their adoption is not widespread yet: to provide a recent example,~\textcite{olszewski2023get} showed that barely half of the publications in top-tier cyber security venues release their artifacts.

We observe that no prior work attempted to analyse the adoption of such repositories for HICSS. Hence, \textbf{we ask ourselves}: ``\textit{To what extent do HICSS papers release their resources into publicly accessible repositories?}'' Such a research question (RQ) serves to understand the overall transparency of HICSS' papers, thereby allowing to: {\small \textit{(i)}}~gauge their replicability and, therefore, the reproducibility of their results; and {\small \textit{(ii)}}~establish a foundation for future work, facilitating methods re-use. 
Moreover, we dissect our main RQ into two sub-RQ by decoupling the term ``HICSS papers'' into two distinct terms:  ``user studies'' (RQ1) and ``technical papers'' (RQ2). Our choice is motivated by the better utility that public repositories have for these classes of papers.

\noindent
\textsc{\textbf{Contributions.}} We want to emphasize the role of external repositories linked within HICSS publications. After defining the problem space in Section~§\ref{sec:background}:
\begin{itemize}[leftmargin=*]
    \item First, through a custom-developed tool, we collect all papers accepted to HICSS in 2017--2024 (§\ref{ssec:collection});
    \item Then, we rigorously review all these works and selectively extract ``user studies,'' entailing human-subject research; and ``technical papers,'' entailing implementation of code or tools (see Fig.~\ref{fig:venn});
    \item Finally, we scrutinize the extent to which such works publicly share their artifacts onto (currently active) external repositories, and inspect their contents (§\ref{ssec:manual}).
\end{itemize}
We present our findings (§\ref{sec:results}) and draw implications for future work (§\ref{sec:discussion}). For the sake of open science, we release our resources at: \url{https://github.com/hihey54/hicss58/}

%% file: sections/2-related.tex
\section{Background and Preliminaries}
\label{sec:background}

\noindent
We position our work within extant literature, and then define our scope and explain its relevance.

\subsection{Related Work}
\label{ssec:related}
\noindent
Abundant prior work has investigated the theme of transparency or reproducibility of research papers. 

Oftentimes, such analyses consider papers published in high-quality venues to provide more compelling findings. For instance,~\textcite{apruzzese2023real} analysed 88 papers published in top-4 security conferences focusing on ``adversarial machine learning'', and found that only half release their source code. Similarly,~\textcite{olszewski2023get} specifically focus on the ``artifacts'' related to top-tier security venues, investigating whether they run as described in the respective publication: out of 298 artifacts, only 20\% produce the same results. 

By adopting a less-technical viewpoint, \textcite{daneshvar2021evidence} consider 82 user surveys within nine top journals in information systems, and found that not one provides enough details for replication. Another perspective is given by~\textcite{miyakawa2020no}, who found that, out of 180 papers submitted to a well-known journal in biology, 41 were explicitly asked to provide raw data before the beginning of the reviewing process: among these, only 1 provided such data. 
The recent study by~\textcite{fivsar2024reproducibility} considered 500 articles in a respected management journal, showing that data is one of 
the most limiting factors in reproducibility. Intriguingly,~\textcite{serra2021nonreplicable} even claim that non-replicable publications are cited more than replicable ones.

We observe that our work also pertains to that branch of research that focuses on how the world embraces  innovations~\parencite{rogers2003diffusion,davis1989technology}. For instance, many studies, such as \textcite{taherdoost2018review}, have investigated how researchers react to progress in the context of information technology. Closely related to our study is the work by~\textcite{escamilla2022rise}, revealing an increased usage of software repositories among academics---according to papers appearing on arXiv (which are not necessarily peer-reviewed) and PubMedCentral (focusing on medicine, which is outside our scope). Nonetheless, in carrying out our analysis, we will borrow the methodological approach (in terms of scripting) adopted in~\textcite{escamilla2022rise}.

To sum up, 
prior work showed that transparency and reproducibility issues may affect a range of domains. Such a phenomenon justifies our choice to focus on HICSS, which is host to cutting-edge and peer-reviewed publications from a constellation of research fields: hence our results can be inspirational for future work.

\subsection{Definitions and Focus of the paper}
\label{ssec:definitions}
\noindent
HICSS papers encompass various types of contributions. We focus on two classes of works: 
\begin{itemize}[leftmargin=*]
    \item \textit{Technical papers}. These articles rely on a strong technical component to test their hypotheses. Examples include: papers that develop an original system/tool \parencite{giudici2024delivering}: papers that use existing tools to carry out some analyses~\parencite{majchrzak2022towards}, potentially with some additional scripts~\parencite{schererhe2022integrated}. Obviously, we do not consider as ``technical'' a paper which relies on trivial software (e.g., a desktop, or a text editor). 

    \item \textit{User studies.} These articles entail human-subject research. Examples include: user surveys~\parencite{sun2021gps}; behavioral studies~\parencite{majchrzak2022towards,yan2023depends}; or interviews~\parencite{nagele2024assessing}.
\end{itemize}
We thoroughly explain the reason for choosing these classes of works in the next subsection. Notably, some papers may fall into both categories: this is typically the case for ``design science research'', wherein interviews or surveys are carried out before developing a certain tool, which is then developed and tested by some users to draw conclusions~\parencite{nguyen2020design}. 

Importantly, some papers may not fall in either category---this is common for literature reviews~\parencite{ul2019business}, or conceptual frameworks~\parencite{asprion2019towards}. The choice to exclude these works is that they have a reduced necessity to upload their resources to repositories (since they typically have no evaluation), and they are hence likely to be self-contained. Therefore, it would be unfair to consider similar works in our analysis, as it would unfairly skew the results---since we seek to analyse HICSS as a whole. 

\subsection{Motivation of our Design Choices}
\label{ssec:statement}

\noindent
The common element of the papers discussed in §\ref{ssec:related} was that they all focused on a single domain. Here, however, we must account for the huge variety of papers (and corresponding methods) that are presented at HICSS. Therefore, to carry out a humanly feasible analysis, we will assess whether any given paper (among the classes we consider) provides the link to an active external repository. Let us explain why this is important.

For technical papers, external repositories are crucial. According to our definition, it is impossible to provide all elements (e.g., code, data, parameter configurations) necessary to accurately reproduce the results in a single manuscript---especially given that HICSS papers are typically 10-pages long without any appendices. Hence, the lack of an external repository may impair the reproducibility of any technical paper.

For user studies, external repositories are also fundamental. For instance, it may be useful if, for interviews, one could release the answers (potentially anonymised). For surveys, sharing the questionnaire could be ideal to inspect how the various questions are presented to the user:  evidence suggests that even the ordering of questions may change the responses of surveys~\parencite{tourangeau2004spacing}. Given that it is unlikely for all such elements to be provided in the paper, existance of repositories is paramount also for this class of works---from a transparency perspective.

The reasons above motivate our choice of focusing on our considered RQs. Moreover, we also investigate \textit{what is included} in the repositories linked in such prior work. Our paper seeks to answer all such questions.

%% file: sections/3-method.tex
\section{Research Method}
\label{sec:method}

\noindent
We describe the methodology followed for our analysis.

\subsection{Data Collection}
\label{ssec:collection}

\noindent
The starting point of our contributions is the collection of HICSS publications spanning 2017--2024; we consider this timespan because HICSS proceedings have been hosted (for free) on \href{https://scholarspace.manoa.hawaii.edu/communities/aaeec9ed-5368-44e3-88e5-33ea62366840}{ScholarSpace} since 2017. We collect such a dataset by developing a custom \textit{scraper}. We provide the code of our scraper in our repository.

Our scraper iteratively downloads HICSS papers from the proceedings web page, organized as follows. From the starting page, it is possible to select a specific edition of HICSS. Then, we are brought to the main tracks of that edition. From here, it is possible to choose the various mini-tracks of the considered track. Finally, upon selecting a mini-track, it is possible to visualize and download the papers included in that mini-track.

We implement our scraper by following the procedure above. Overall, we downloaded a total of 5579 publications (whose cumulative size is $\approx$5.42GB). Notably, this number matches the one provided by summing all the numbers of accepted papers of HICSS as mentioned in their preface (i.e., 5579). This confirms the quality of our scraper, and the soundness of our methods. The resulting dataset has the same organization described above (years-tracks-minitracks-papers) to facilitate navigation.

\textbox{{\small \textbf{Ethics:} We cannot publicly release our dataset. However, by using our scraper, one can collect the same dataset: the HICSS' proceedings are publicly available, free of charge.}}

\vspace{-6mm}

\subsection{Preliminary (automatic) Paper Analysis}
\label{ssec:preliminary}

\noindent
The second step entails analysing the 5579 collected publications. However, given their large number, we carry out a preliminary analysis in an automated way by developing a \textit{custom script} (provided in our repository), as also done by~\textcite{escamilla2022rise}.

At its core, our script iteratively inspects the content of each paper (a .pdf file) in our dataset. For every paper, the script is designed to answer three questions: ``does the paper entail a user study?'', ``does the paper involve some technical implementation?'', ``does the paper provide a link to a dedicated repository?''. To answer each of these questions, the script looks for certain keywords included in a set of \textbf{three lists} which we autonomously devised. Specifically:
\begin{itemize}[leftmargin=*]
    \item \textit{user study}: variations of {\small (user study, questionnaire, online survey, user survey, interview)};
    \item \textit{technical paper}: variation of {\small (empirical, experiment, evaluation, source code, artifact, implementation, tool)};
    \item \textit{repositories}: variations of {\small (github.com, gitlab.com, zenodo.org, figshare.com, anonymous.4open.science, drive.google.com, onedrive.live.com, 1drv.ms)}.
\end{itemize}
The exact lists, and corresponding patterns, are observable in the code of our script. We derived these lists after having qualitatively inspected some papers, but also through our own domain expertise. 

Hence, we let the script run and process all papers included in our dataset. To \textbf{identify potential candidates} for the ``user study'' or ``technical paper'' categories, the respective lists of keywords are checked \textit{only in the abstract}: if any match is found, the paper is considered as a candidate for the corresponding category. We only check the abstract because our results would be too noisy and unreliable if we considered the whole paper. Importantly: the search does not stop at the first occurrence. Therefore, the script is also likely to assign the same paper to both categories. 
Simultaneously, the script also checks the \textit{entire content of the paper} for patterns mimicking the corresponding list that could point to external resource repositories.

At the end of this preliminary procedure, we obtain a set of 1949 and 551 papers: the former being potential candidates for technical papers (of which 133 may have a dedicated repository), the latter for user studies (of which 11 may have a dedicated repository); 173 papers are placed in both categories. 3425 papers are not included in any category, and will not be considered in the following analysis. 

\begin{figure*}[!htbp]
    \centering
    \includegraphics[width=2\columnwidth]{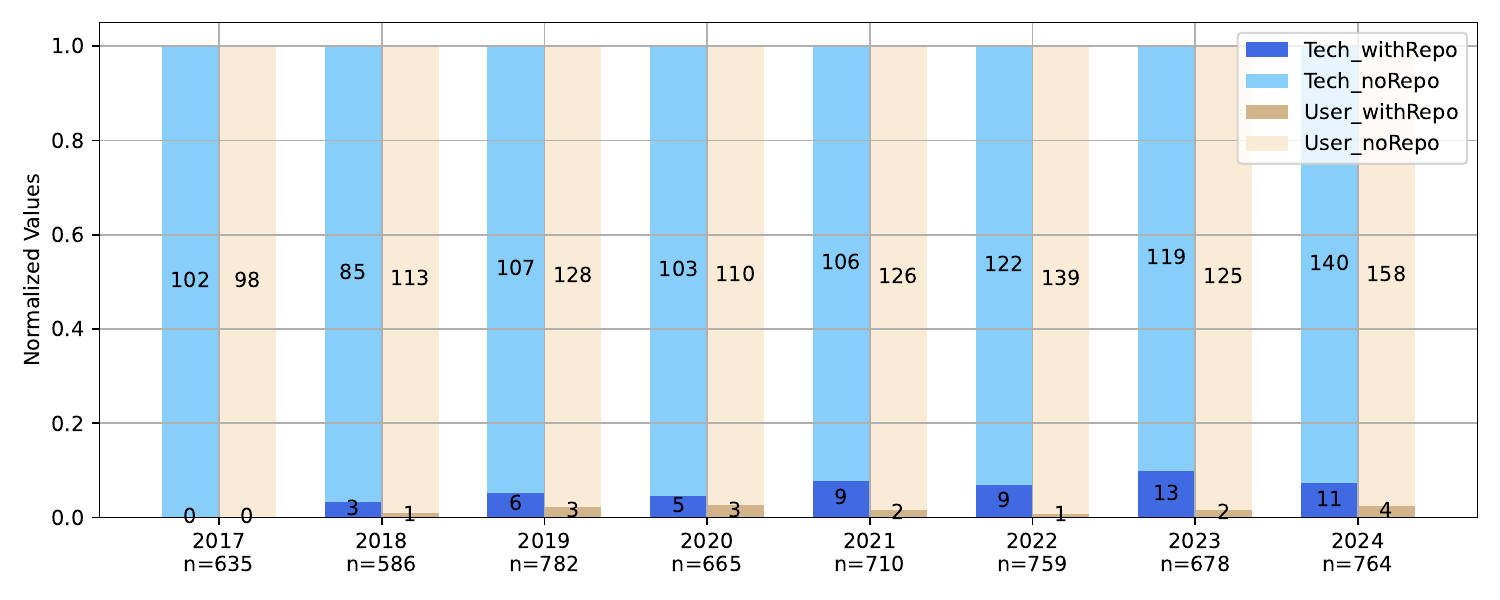}
    \vspace{-0.5cm}
    \caption{{\small \textbf{Temporal distribution of our analysed papers}} --
    \textmd{\footnotesize The numbers inside the bars are the absolute numbers, while the bars represent the proportion of the papers (user studies and technical) that have (or not) functional repositories. Ticks on the x-axis report the year and the number of papers accepted to HICSS for that year.}} 
    \label{fig:temporal}
    \vspace{-3mm}
\end{figure*}

\subsection{Manual Review and Validation}
\label{ssec:manual}

\noindent
It is unrealistic to expect that a simple and fully-automatic keyword check guarantees accurate extraction of the information we need to answer our RQs. Hence, as the last step of our procedure, we manually review the papers in each group of candidates. The intention is threefold: {\small \textit{(i)}}~ensure that the papers are included in the correct category; {\small \textit{(ii)}}~verify the paper has a link to its specific repository, and {\small \textit{(iii)}}~check the content of the repository---if available.

Such an analysis is \textbf{done in pairs}, in a similar fashion as done by~\textcite{apruzzese2023real}: two authors work independently to inspect the papers and assign the category to each of them. The authors frequently discussed their findings, and if there were any uncertainties about their judgements, they consulted each other to reach a consensus.\footnote{During the reviewing phase of this paper (hence \textit{after} we had carried out our analysis) we have carried out an experiment to measure the \textbf{inter-coder reliability score}. Each author randomly chose 50 papers that they independently reviewed, and asked the other author to assign the corresponding class. We observed an agreeability of 93\%, denoting that the authors likely reach the same conclusion.} These analyses are qualitative in nature, but follow a rigorous approach encompassing the entire text of the paper. For instance, for ``user study'', we used manual keyword search (with similar terms as in the corresponding list), but also visual inspection (e.g., tables or figures); for ``technical papers'', we examine if non-trivial software\footnote{We consider a ``trivial software'' as those programs that are a prerequisite to carry out research activities, e.g., MS Word for writing, or Excel for analytics or visualizations, or PowerPoint for slides.} has been used during the study---either by using existing tools with minimal changes, or modifying previous scripts, or developing new systems; whereas for ``repositories'', we check if the links point to a repository that is truly associated to the activities carried out in the paper (e.g., links pointing to repositories created beforehand, or from different authors, are not counted). 

To provide more insightful results, we not only scrutinize if a paper is a user study or a technical paper, but \textbf{we delve deeper}. 
For technical papers, we are particularly interested in the nature of the provided repositories, asking ourselves ``do these repositories include data, code, or an entire system?''. For user studies, we first check if the paper is an observational study, or a user survey, or interviews; and then explore their repositories, checking whether they include supplementary material that is related to the nature of the user study (e.g., questionnaires). Hence, we manually check the content of all repositories associated with any given paper included in our analysis.

We show a summary representation of the results of our manual analysis in Fig.~\ref{fig:venn}. Before addressing our RQs, we find it instructive to determine the \textbf{quality of our custom script}. We do so by measuring how many papers, after our manual inspection, had been correctly classified by our script. For user studies, our script identified 551 papers, but our manual check found 997 (some of these were identified in the set of technical papers); for technical papers, our script identified 1949 papers, but we found 884; for matching repositories, our script identified 144, but we found 71. Hence, we endorse downstream researchers to \textit{exercise caution} when using our script to perform similar analyses, as the number of misclassifications is substantial.

%% file: sections/4-results.tex
\section{Results}
\label{sec:results}

\noindent
We now present the results of our research after our manual validation.
To this end, we first provide in Fig.~\ref{fig:temporal} the temporal distribution over the considered timespan (2017--2024) of our findings. This plot shows how many papers for each class have a functional link\footnote{By ``functional link'' we mean that it leads to a repository that is active and has some content, and hence serves a ``function''.} to a corresponding external repository (papers falling in both classes are counted in each class). Specifically, the x-axis shows the years, and for every year the number of total papers accepted at HICSS; the y-axis reports the relative distribution of the papers with/without an external repository; numbers on the bars denote the absolute number of papers. We can already see that the overall percentage of papers with repositories is abysmally low ($\approx$3\%). However, let us further analyse our findings by focusing on our RQs.

\subsection{User Studies (RQ1)}
\label{ssec:userstudy}

\noindent
We first focus on papers entailing user studies.

First, out of 997 papers entailing user studies, only 16 (1.6\%) provide supplementary resources in an external repository. According to Fig.~\ref{fig:temporal}, it is evident that this number ranges from 0 to 4 for each of the eight years. In the year 2017, there were no user studies that linked external repositories. However, going forward it is visible that this number started to increase. As a result, in the years 2019 and 2020, 3 studies in each year provided their supplementary materials in linked repositories. In this regard, the last year (2024) had the highest number (4), which can be considered encouraging and could indicate that the growing trend will continue. Moreover, it is also noticeable that the number of user studies tends to increase every year (also w.r.t. HICSS overall papers): 98 papers (out of 635) included user studies in 2017, 129 (out of 665) in 2021, 140 (out of 759) in 2022 and 162 (out of 764) in 2024. 

Second, delving deeper, we found that, out of 997 papers having user studies, 472 (47\%) entail surveys, whereas 595 (60\%) interviews and only 42 (4\%) are behavioral studies.\footnote{We note that the sum of these percentages exceeds 100\% because multiple papers contained ``combinations'' of surveys and interviews (sometimes even behavioral studies).} In regards to user surveys, 11 studies (2\%) provide the URL to the external repository. The numbers do not differ extensively for interviews (7, i.e., 1\%) and behavioral studies (1, i.e., 2\%). This is presented in Table~\ref{tab:user}.

Third, focusing on the contents of the provided repositories, we found that accessing repositories can also be a problem. Out of 16 repositories, we were able to access 13 (81\%) repositories. There are two reasons for this: either the \textit{URL is broken}, or the \textit{repository is private} (we do know if these issues are only occurring at the time of our analysis in June 2024, or have always been present). This can make it difficult to evaluate any shared contents as well as to reproduce the conducted studies. Furthermore, we checked the contents of the 13 accessible repositories. The shared materials included various artifacts such as surveys, interview descriptions, codebooks, questions, data, results, code, and tools. It is important to note that the last two mentions relate to user studies that could be also considered technical papers.    

Finally, we found that user studies represent 19\% of the overall number of HICSS publications, emphasizing the role of user-centered research in the field of information systems and technology. This is further underscored by the fact that it is growing every year, indicating a sustained commitment to studies involving human subject research. We also found that a small percentage (2.6\%) of user studies can be classified also as ``technical papers'', a classification that is particularly popular for some more technical tracks at HICSS, such as \textit{Software Technology}. This indicates that a subset of user studies delves into the technical aspects of software development, implementation, or evaluation, showcasing the intersection of user-focused research with technical domains.

\begin{cooltextbox}
\textsc{\textbf{Takeaway (RQ1).}} The percentage of user studies in HICSS publications is significant at 19\%, but the provision of supplementary resources in publicly available external repositories remains low at 1.7\%. 
\end{cooltextbox}

\begin{table}[!htbp]
    \centering
    \caption{\textbf{Distribution of papers with user studies --}
    \textmd{\footnotesize 
    We show the number of the subclasses of user studies that have functional repositories.}} 
    \label{tab:user}
    \vspace{-2mm}
    \resizebox{0.75\columnwidth}{!}{
        \begin{tabular}{c?c|c}
            \toprule
            
            Subclass & Total & Repo?  \\
            \midrule
            Users Surveys & 472 & 11 \\
            Users Interviews & 595 & 7 \\
            Behavioral Studies & 42 & 1 \\

            \bottomrule
        \end{tabular}
    }
\end{table}

\subsection{Technical Papers (RQ2)}
\label{ssec:technical}

\noindent
To provide answers to RQ2, we discuss technical paper contributions.

Out of 884 papers with a substantial technical component, only 56 (6\%) provide supplementary resources in a publicly available external repository. As presented in Fig.~\ref{fig:temporal}, the trend of including supplementary resources in an external repository is generally increasing. In 2017, no papers provided an external repository, but this number rose to 6 in 2019, 9 in both 2021 and 2022, and reached 13 in 2023. However, there was a slight decrease to 11 in the year 2024, indicating a fluctuation in the provision of supplementary resources over the years. This trend demonstrates a growing commitment to sharing resources, despite minor variations in the numbers from year to year---a finding which aligns with those by~\textcite{apruzzese2023real}.

\begin{table}[!htbp]
    \centering
    \caption{\textbf{Contents of the repositories for technical papers -- }
    \textmd{\footnotesize 
    We show the most prevalent contents after analysing these repositories.}} 
    \label{tab:technical}
    \vspace{-2mm}
    \resizebox{0.5\columnwidth}{!}{
        \begin{tabular}{c?c}
            \toprule
            
            Content type & Count \\
            \midrule
            Source Code & 33\\
            Data & 12\\
            Tool & 16\\
            Other & 9\\

            \bottomrule
        \end{tabular}
    }
\end{table}

Focusing on the contents of the provided repositories, we found out that these usually contain source code, data, or tools. However, we also added the ``other'' category for any content that does not fit in any of the aforementioned ones. This is usually related to documentation, knowledge bases, and shared content with user studies. According to our investigations (see Table~\ref{tab:technical}), source code (33, i.e., 59\%) is more often shared compared to the other content types. The code is usually related to a certain implementation of an algorithm or a script. This is followed by tool\footnote{Tools can be considered as more mature implementations compared to the source-code. For instance,~\textcite{mclean2023design} provide a tool, whereas only some code snippets and~\textcite{salminen2021automatically}. Note that neither is better or worse than the other.} (16, i.e., 29\%) and data (12 i.e., 21\%) repositories. Finally, the ``other'' category was identified 9 (16\%) times. Of course, some repositories may have contained more than one of these types of content.

Intriguingly, we found that in 4 (7\%) cases, the provided link was not functional, or was not publicly available, or it pointed to an empty repository or a repository was mentioned but with no link (e.g., ``code available on GitHub'' but without any link). According to this, it can be presumed that in most of cases, the shared repositories are functional and accessible.

Finally, we found that, in general, technical papers represent 16\% of the overall number of HICSS publications, which can be considered a significant number for an information systems conference. We also note that, intriguingly, technical papers were more popular than user studies in 2017 (102 vs 98), but this occurrence changes starting from 2018 (88 vs 116): since then, technical papers have always been outnumbered by papers with user studies (e.g., 151 vs 154 in 2024).

\begin{cooltextbox}
\textsc{\textbf{Takeaway (RQ2).}} Only 6\% out of 884 technical papers have a link to a publicly available repository. This percentage is superior to user studies, but external repositories are also more valuable for technical papers in the context of reproducibility and usefulness for downstream research (e.g., code reuse).
\end{cooltextbox}

\subsection{Tracks Distribution (ancillary RQ)}
\label{ssec:distribution}

\noindent
We conclude by considering an ancillary research question: ``which \textit{tracks} accept most papers that include a functional link to a corresponding repository?'' 

To answer such a research question, we show in Table~\ref{tab:tracks}, the overall distribution of the papers across the 13 most ``stable'' tracks\footnote{Every year, HICSS may have new tracks. We only consider the tracks that have always appeared at HICSS in our considered timespan.} of HICSS during our considered timespan (2017--2024) as well as the percentage of those that provide (or not) an external repository. For instance, for the \textit{Software Technology} track and the \textit{Software, Engineering, Education and Training} track, the overall percentage of papers with an external repository is 9.6\% and 6.5\%. These results suggest that these tracks are likely to provide more supporting materials to replicate the studies or to gain additional insights into the materials that are not part of the original publication. 

In contrast, the \textit{Internet Digital Economy} track and the \textit{Information Technology in Healthcare} track have an overall percentage of papers with an external repository of 1.4\% and 1.9\%. This indicates that despite the relatively high number of studies conducted in these research areas, there are fewer supporting materials provided. The tracks as the \textit{Internet Digital Economy}, \textit{Collaboration Systems and Technologies}, \textit{Information Technology in Healthcare}, and the \textit{Software, Engineering, Education and Training} track had the highest number of studies, yet relatively lower overall percentage against the repositories. 

Looking at the categories of technical papers individually, the tracks \textit{Software, Engineering, Education and Training} as well as \textit{Software Technology} had a percentage of technical papers with an external repository of 11.8\% and 11.7\%. An example of a recent technical paper with a repository from this particular track is the paper by~\cite{pekaric2024attack}. As for the lowest percentage, the \textit{Information Technology in Healthcare} track had a total of 85 papers, yet only 1.1\% contained an external repository. With that said, the overall percentages of the technical papers still ranged from 7.3\% to 11.8\%. This suggests that the majority of the topics provided a relatively low number of supporting materials---at least compared to papers of highly technical venues, such as those investigated by~\textcite{olszewski2023get}.

The tracks that contained user study papers, on the other hand, resulted in fewer papers providing links to functional external repositories, with the highest having a percentage of 4.3\% only (\textit{Software Technology} track). Track \textit{Decisions, Analytics, Services, and Science} had the lowest percentage of 0.09\%. Track \textit{Collaboration Systems and Technologies} was also not far behind at 1.3\%. This indicates that the user studies offered fewer supporting materials than the technical papers. 

\begin{table}[!htbp]
    \centering
    \caption{\textbf{Distribution of papers over the tracks --}
    \textmd{\footnotesize 
    We show the percentage of papers for each ``stable'' track of HICSS.}} 
    \label{tab:tracks}
    \vspace{-2mm}
    \resizebox{\columnwidth}{!}{
        \begin{tabular}{c?c|c|c|c}
            \toprule
            \multirow{2}{*}{\textbf{Track Name}} & \multicolumn{2}{c|}{\textit{Technical Papers}} & \multicolumn{2}{c}{\textit{User Study}} \\ \cline{2-5}
            & Total & Repo? & Total & Repo? \\
            \midrule
            Decis. Analyt. Serv. Scie. & 141 & 10 & 103 & 1 \\
            Digital Social Media & 64 & 5 & 83 & 1\\
            Electric Energy Systems & 55 & 4 & 1 & 0\\
            Digital Government & 58 & 3 & 90 & 2\\
            Internet Digital Economy & 90 & 1 & 130 & 2 \\
            Know. Innov. Entr. Syst. & 29 & 2 & 69 & 2\\
            Software Technology & 140 & 16 & 47 & 2\\
            Location Intelligence & 20 & 0 & 4 & 0\\
            Inf. Tech. Social Just. & 9 & 1 & 8 & 0\\
            Collab. Syst. Techn. & 126 & 6 & 151 & 2\\
            Inf. Tech. Healthcare & 85 & 2 & 123 & 2\\
            Soft. Eng. Educ. Train. & 17 & 2 & 14 & 0\\
            Org. Syst. Techn. & 50 & 4 & 174 & 2\\

            \bottomrule
        \end{tabular}
    }
\end{table}

%% file: sections/5-discussion.tex
\section{Discussion and Lessons Learned}
\label{sec:discussion}

\noindent
We discuss our findings, outlining limitations and then derive recommendations for future work.

\subsection{Threats to validity and Mitigations}
\label{ssec:limitations}
\noindent
We identify two limitations of our analysis.

First, our results stem from a manual analysis carried out after the application of our preliminary filtering script (see §\ref{ssec:preliminary}). The script is built on a keyword search of the abstracts of HICSS papers. As such, \textit{it is possible that our results underestimate} the overall number of user studies or technical papers: some papers may very well discuss research that could fall into our definitions (§\ref{ssec:definitions}), but which have not been included in our manual analysis due to their abstract not including our considered search terms. The same applies to repositories: we considered keywords of well-known platforms, but if a paper provided their resource on a custom website, our script may not identify it.\footnote{However, we may still ``catch'' these cases \textit{if} the text included the other keywords. An example is the work by~\textcite{milliken2021behavioral}.} 

Second, our validatory analysis has been carried out qualitatively by two authors, who manually reviewed thousands of papers. It is possible that, during this process, some papers have been assigned to the wrong category, or some links have been overlooked. Due to a lack of ground truth, it is impossible to guarantee ``perfect'' results. This is why we make our results fully available for scrutiny by fellow researchers: \textbf{in our repository, we have included a table showing the breakdown of our analyzed papers}, specifying the category, the link to the repository we identified, and the contents we found in such a repository.\footnote{We will update the table if researchers provide factual reasons.}

A potential workaround to ``automatically'' address the abovementioned shortcomings is to rely on large language models (LLM). For instance, by submitting our papers to some LLM-based tools\footnote{As an example in 2024: \url{https://www.chatpdf.com/}} and asking ``does the paper entail human-subject research?'' or ``does the paper include a repository that is specific of the research done in the paper?'' it may be possible to derive more accurate results. However, despite their potential, LLMs have explainability issues~\parencite{zhao2024explainability}. This is why we preferred to carry out our study ``by ourselves,'' instead of relying on AI. Nevertheless, given that we publicly release our resources, future work can compare our results with those produced by various LLMs.

\subsection{Implications and Disclaimers}
\label{ssec:implications}

\noindent
To avoid generating harmful misunderstandings, let us clarify the implications of our findings.

First, it is \textbf{wrong to conclude} that the overall ``lack of external repositories'' is a sign that prior work is unreliable or,  worse, of poor scientific value. Indeed, there are plenty of reasons why such supplementary material is not provided. For instance, in the case of source code, the owners may not be allowed to share it due to, e.g., industrial secrets. In the case of data, there may be privacy-related reasons.\footnote{It may even be that the size of the data is so big that it cannot be included in a single repository due to limitations~\parencite{freeman2022workflow}, and the data can hence only be provided upon request.} It may even be entirely possible that the paper provides all the necessary elements to replicate the same research---and, hence, there is no need of a third-party repository. Furthermore, additional material for replication can be obtained by contacting the authors. Moreover, for full-fledged systems, it is legitimate to avoid sharing their implementation if the underlying code is not ready for public release: maybe the code is poorly commented, or has no documentation, and publicly releasing such ``impractical'' code would not be of much use. Finally, HICSS allows to attach ``supplementary material'' in the submission system: it is possible that the reviewers were provided with extra resources which have been scrutinized, but not released due to any of the abovementioned reasons---thereby further validating the work described in the paper. Therefore, \textit{we refrain from carrying out analyses focusing on individual papers}---as done, e.g., by~\textcite{konigstorfer2024black, olszewski2023get}

Nonetheless, it is a fact that only a tiny percentage of papers accepted at HICSS provides publicly observable supplementary resources. We believe that the ``magnitude'' of the research discussed in each paper accepted to HICSS is far greater than the content of the paper itself. Indeed, to quote the overarching message of the ``artifact evaluation'' process of top-tier security conferences\footnote{A list can be found at: \url{https://secartifacts.github.io/}.} ``\textit{Your paper is more than just words. Its artifacts extend beyond the document itself: software, hardware, evaluation data and documentation, raw survey results, mechanized proofs, models, test suites, benchmarks, and so on. In some cases, the quality of these artifacts allows for easy extension and reproduction.}'' Hence, we endorse the organizers of HICSS' tracks and respective mini-tracks to promote the release of additional resources, and/or to consider in higher regard those papers that provide such artifacts. Potentially, submitted papers that include artifacts can be marked with specific labels to clearly distinguish them. Finally, we emphasize that, during our manual review, it was not simple to determine if a paper with a dedicated repository actually \textit{had} such a repository, since the link was often hidden among a myriad of other links. Hence, the link to such an artifact could be added to the page of the accepted paper.

\subsection{Directions for Future Work}
\label{ssec:future}

\noindent
The research discussed in this paper provides a stepping stone for future work---not necessarily pertaining to the field of transparency/reproducibility.

First, our supplementary material includes \textit{the links to all repositories we found} during our analysis. Hence, future work can use such links as a guide to, e.g., build new tools, investigate ``why'' some authors did (or did not) provide their resources in a publicly available repository, or carry out novel research by relying on the artifacts released by prior publications---thereby further increasing the impact of HICSS' prior contributions. Alternatively, future work can carry out true reproducibility studies by using the repositories we found and assessing their functionality---as done, e.g., by~\textcite{olszewski2023get}. Such analyses can be beneficial as well as inspirational for those fields of research for which transparency and reproducibility are known to be of crucial importance, such as AI~\parencite{gesing2023science}.

Another avenue is \textit{expanding our findings}: we release all our resources, so future work can use them to investigate if upcoming editions of HICSS will reflect the same trend as those covered in this paper; or even if editions of HICSS that occurred before our considered timespan (e.g., before 2017) have a different trend. At the same time, it is possible to use our dataset (which can be recreated by using our scraper §\ref{ssec:collection}) to carry out orthogonal analyses focused on the overall evolution of the publications of HICSS. For instance, an intriguing phenomenon that we noticed during our manual review is the ``different formatting'' that HICSS publications tend to have: this can happen either during different editions (e.g., compare~\cite{sedlmeir2021dlps} with~\cite{koh2024voices}), or even across the same edition of HICSS (e.g., compare~\cite{rieger2018evaluating} with~\cite{gabbrielli2018language} and~\cite{scrivner2018topic}).

An intriguing direction is to find \textit{ways to improve the research methods used in our work}. For instance, one can use large-language models and see if such models can produce similar results to ours in a fraction of the time (it took us weeks to carry out all our manual verifications!). Moreover, it is also possible to improve our script (which is provided in our repository). To this end, we provide two lessons learned from our analysis: {\small \textit{(i)}}~keywords can be expanded to include additional synonyms and domain-specific terms; {\small \textit{(ii)}}~regular expressions can be utilized for more flexible pattern matching. Alternatively, machine learning approaches, such as topic modeling, can be used to remove hard-coded keyword searches and increase flexibility---potentially allowing to identify papers that we missed, but at the expense of some ``false positives''.

Finally, future work can also \textit{investigate the papers excluded from our analysis} (see §\ref{ssec:definitions}). For instance, some literature reviews can be carried out by means of AI, and may hence include a strong technical component. Whereas some works that are neither technical in nature nor include human-subject research could provide supplementary resources that expand their primary contributions. It is hence intriguing to explore this additional dimension, which would shed more light on HICSS' history of publications.

%% file: sections/6-conclusions.tex
\section{Conclusions}
\label{sec:conclusions}

\noindent
We have investigated the extent to which HICSS publications within 2017--2024 release additional resources in external repositories. Our analysis, carried out with a mix of scripting and manual reviewing, revealed that only 3\% of the 2028 papers that we considered have an external link to an external repository containing supplementary material.

Our efforts can serve as an inspiration for future work: although papers have a limited number of pages, external repositories are not bound by such limits. Such repositories can hence be used to share the dearth of artifacts that are generated in the process of deriving the results of any given research paper. We invite future work to reflect upon our findings: providing supplementary resources not only facilitates downstream research, but also increases the soundness of each accepted paper---thereby increasing their overall contribution to the state of the art.